\documentclass{article}
\usepackage{spconf,amsmath,graphicx,hyperref}
\usepackage{booktabs}
\usepackage{amsfonts}
\usepackage{subfig}
\usepackage{bm} 
\usepackage{xcolor} 
\title{Gen-SER: When the generative model meets speech emotion recognition}
\name{Taihui Wang$^{1}$$^{2}$, Jinzheng Zhao$^{1}$$^{2}$$^{4}$, Rilin Chen$^{1}$$^{2}$, Tong Lei$^{2}$, Wenwu Wang$^{4}$, Dong Yu$^{3}$}
\address{$^{1}$Tencent Multimodal Models Department,
Beijing, China\\
$^{2}$Tencent AI Lab, Beijing, China\\
$^{3}$Tencent AI Lab, Bellevue, WA, USA\\
$^{4}$Centre for Vision, Speech and Signal Processing, University of Surrey, UK}
\begin{document}
%\ninept
%
\maketitle
\begin{abstract}
Speech emotion recognition (SER) is crucial in speech understanding and generation. Most approaches are based on either classification models or large language models. Different from previous methods, we propose Gen-SER, a novel approach that reformulates SER as a distribution shift problem via generative models. 
We propose to project discrete class labels into a continuous space, and obtain the terminal distribution via sinusoidal taxonomy encoding. The target-matching-based generative model is adopted to transform the initial distribution into the terminal distribution efficiently. The classification is achieved by calculating the similarity of the generated terminal distribution and ground truth terminal distribution.
The experimental results confirm the efficacy of the proposed method, demonstrating its extensibility to various speech-understanding tasks and suggesting its potential applicability to a broader range of classification tasks.
\end{abstract}
\section{Introduction}
\label{sec:intro}

Speech emotion recognition (SER) is an important task in speech understanding and is beneficial for human-computer interaction \cite{pandey2019deep}. It has wide applications in quality assessment \cite{huahu2010application} and monitoring \cite{lieskovska2021review}, and can also be used as a reward model for the text-to-speech system \cite{du2025cosyvoice}. SER can also be included as an auxiliary task for the speech tokenizer to facilitate learning of semantic information \cite{gao2025differentiable}.

% \section{Related work}
% \label{sec:related work}
% In this section, we provide a concise review of two widely used algorithm of SER, namely the classification model and LLM for SER. In addition, several generative models for the classification task are discussed.
% \subsection{Classification model}
Previous methods typically employ a classification pipeline consisting of a speech encoder followed by a classifier.   Early work used processed representations such as the short-time Fourier transform  \cite{mustaqeem2019cnn} 
% raw audio \cite{zhao2019speech} or
and Mel spectrogram \cite{pandey2019deep} as input features. Neural modules like convolutional neural network (CNN) \cite{mustaqeem2019cnn} or long short term memory (LSTM) \cite{pandey2019deep} were then applied to process these features, with a fully connected layer serving as the classifier.  
% or SVM \cite{huang2014speech}
% These methods suffer from poor generalization and a heavy reliance on large amounts of labeled training data.
More recently, pretrained models have been adopted for feature extraction:   for example, 
% Vesper \cite{chen2024vesper} is a pretrained model based on WavLM \cite{chen2022WavLM} for SER leveraging emotion-guided masking and hierarchical self-supervision. Furthermore, 
emotion2vec \cite{ma2023emotion2vec} 
% was initialized from data2vec \cite{baevski2022data2vec} and
leverages online knowledge distillation combined with supervised fine-tuning for emotion recognition. 
% DropFormer \cite{mai2024dropformer} uses WavLM \cite{chen2022WavLM} for feature extraction and leverages the attention mechanism to attach importance on emotional information while discard non-emotional segments. There are also works incorporating paralinguistic information like gender and age \cite{wan2025metadata} to enhance the emotion recognition performance. 
% \subsection{LLM for SER}
With advances in large language models (LLMs),  several studies \cite{chu2023qwen, chu2024qwen2} have investigated LLM-based approaches for SER.   In such pipelines, a speech encoder (e.g.,  
% Whisper \cite{radford2023robust} or
WavLM \cite{chen2022WavLM}) extracts speech representations that are aligned to the LLM input via an adapter, and the LLM is prompted (e.g., “Describe the emotion of the speaker”) to produce the emotion label.  
% Although these methods have achieved impressive results, they face relatively high training and inference costs.
% Qwen-audio \cite{chu2023qwen} adopts a multitask input format in which each tag denotes a speech-understanding task, and Qwen2-Audio \cite{chu2024qwen2} further incorporates DPO \cite{rafailov2023direct} to improve understanding. 
% In \cite{zhang2025soundwave}, Soundwave is proposed, which uses CTC \cite{graves2006connectionist} loss to reduce the length of the speech features and map the vocabulary space from the encoder to the LLM. 
% There are also methods using dual encoders for feature representation. Kimi-audio employs semantic tokens from GLM-4-Voice and add the features from whisper encoder with vector quantization. 

% \subsection{Generative model for classification}
Li et al. \cite{liYourDiffusionModel2023} propose a diffusion-based classifier  that treats classification as a conditional density-estimation problem, where class labels serve as conditional inputs, and classification is achieved by selecting the class that minimizes the noise-prediction error. 
This approach bridges the gap between generative and discriminative paradigms. However, it suffers from computational inefficiency, as inference latency scales linearly with the number of classes due to per-class error estimation requirements. 
Another related work applies generative modeling to voice activity detection (VAD) by projecting binary VAD labels into a latent space via an autoencoder \cite{chenFlowTSVADTargetSpeakerVoice2025}.
Compared with the per-class evaluation required by the diffusion-based classifier \cite{liYourDiffusionModel2023}, latent-space projection can improve computational efficiency by avoiding separate processing for each label.
% Although latent space projection enables efficient flow matching, its reliance on reconstruction fidelity poses a bottleneck for scalable deployment, since training an auto-encoder for multi-class scenarios is non-trivial and inevitably introduces reconstruction errors during the label space transformation.
While latent-space projection enables efficient flow matching, its reliance on reconstruction fidelity hinders scalable deployment as training autoencoders for multi-class tasks is challenging and inevitably causes reconstruction errors during label space transformation.

In this paper, we propose Gen-SER, a generative approach that reframes SER as a distribution-transport problem. Our contributions are threefold. First, we are the first to reformulate SER as a distribution transport problem via generative models, to our knowledge. Second, we propose the sinusoidal taxonomy encoding method and map discrete labels to continuous space, avoiding multi-class autoencoder training hurdles. Third, our proposed target-matching generative model with logistic mean and bridge variance schedules enables efficient distribution transport. 
% Experimental results validate the efficacy of the proposed approach in SER. Moreover, the results in the gender classification task demonstrate the extensibility of the proposed method, suggesting its potential application in other classification tasks.
Experimental results demonstrate the efficacy of the proposed method in SER. Furthermore, its performance on the gender classification task highlights the model's extensibility, indicating considerable potential for adaptation to a broader range of classification tasks.

\section{The Proposed Method}
\subsection{Methodology}
We assume: 1) speech signals expressing different emotions follow distinct initial distributions; 2) each class label corresponds to a pre-defined terminal distribution; and 3) distributions can be transported from the initial to the terminal distribution. Based on these assumptions, the SER task can be formulated as a distribution transport problem. Given an observed data sample $\mathbf{x}_1$ drawn from an emotion-specific distribution, we transport it to $\mathbf{x}_0$ that follows the target distribution of a class via a generative model. Finally, we compute the similarity between the generated $\mathbf{x}_0$ and each class’s target distribution, assigning the sample to the class with maximal similarity.

\subsection{Generate the data sample $\mathbf{x}_1$ using the speech signal}
Unlike conventional generative models that sample from a Gaussian prior, we use the pre-trained HuBERT model \cite{hsuHuBERTSelfSupervisedSpeech2021} to extract meaningful features from the speech signal. We use output from the final HuBERT layer as $\mathbf{x}_1$ for the input and use output from the preceding layers as conditioning $\mathbf{X}_c$.
\begin{equation}
	\mathbf{X}_c,\mathbf{x}_1 = Average(HuBERT(\mathbf{s})),
    \label{eq: ssl}
\end{equation}
% \begin{equation}\label{eq: ssl}
% 	\mathbf{X}_c = [\mathbf{h}_0,\mathbf{h}_1,\cdots,\mathbf{h}_{K-2}]
% \end{equation}
where $\mathbf{s}$ is the time-series speech signal, $HuBERT(\cdot)$ denotes processing the speech signal by the pre-trained HuBERT model, and $Average(\cdot)$ means averaging along the temporal axis. 
% Note that $\mathbf{x}_1$ and $\mathbf{h}_i$ are embedding vectors with length of $L$.

\subsection{Generate the data sample $\mathbf{x}_0$ using the class label}
Instead of training an auto-encoder to map class indices into embedding vectors \cite{chenFlowTSVADTargetSpeakerVoice2025}, we first map the distinct categories to their corresponding ordinal indices, and then generate  embedding vectors in the continuous space based on these indices using  sinusoidal taxonomy encoding method
\begin{equation}\label{eq: sin}
	\mathbf{x}_0(b) = sin \left( \frac{2 \pi \bm{l}}{L}*(i_b+1) \right),
\end{equation}
where $b$ denotes the class label, $i_b$ denotes the index of class $b$, $L$ is the \textcolor{black}{length} of the embedding vector, and $\bm{l}$ denotes a vector consisting of integers ranging from 0 to $L-1$. This encoding method has two advantages. Firstly, the generated embedding vectors are continuous, which facilitates learning. Secondly, the embedding vectors of distinct categories are mutually orthogonal.
% \begin{equation}\label{eq: orthogonal}
% \langle \mathbf{x}_0(m), \mathbf{x}_0(n) \rangle = 
% \begin{cases} 
% 1 & \text{if } m = n, \\
% 0 & \text{if } m \neq n,
% \end{cases}
% \end{equation}
% where $ <\cdot,\cdot>$ denotes the inner product between two vectors. 
In the following discussion, the class label $b$ in $\mathbf{x}_0(b)$ is omitted for conciseness unless stated otherwise.

\subsection{Generative model for distribution transport}
The goal of the generative model is to generate $\mathbf{x}_0$ given $\mathbf{x}_1$ step by step. 
This distribution transport problem is modeled by an ordinary differential equation (ODE) \cite{lipmanFlowMatchingGenerative2023}
\begin{equation}\label{eq:ode}
	\frac{d \mathbf{x}_{t}}{dt}=\bm{u}\left(\mathbf{x}_{0}, t,\mathbf{x}_{1}\right),
\end{equation}
where the vector field $\bm{u}\left(\mathbf{x}_{0}, t,\mathbf{x}_{1}\right)$ determines the evolution of the perturbed signal $\mathbf{x}_{t}$, guiding the flow along the probability path.
Considering a special case where $\mathbf{x}_{t}$ is Gaussian
\begin{equation}
	p(\mathbf{x}_t) = \mathcal{N}(\mathbf{x}_t;  \textcolor{black}{\bm{\mu}_{t} (\mathbf{x}_0,\mathbf{x}_1)}, \sigma^2 (t)\mathbf{I}),
\end{equation}
where $\mathbf{I}$ denotes the identity matrix, the vector fields can be obtained according to \cite{lipmanFlowMatchingGenerative2023} as
\begin{equation}\label{eq: vetor field}
	\bm{u}\left(\mathbf{x}_{0}, t, \mathbf{x}_{1}\right)=\frac{\sigma_{t}^{\prime}}{\sigma_{t}}\left(\mathbf{x}_{t}-\textcolor{black}{\bm{\mu}_{t}\left( \mathbf{x}_{0},\mathbf{x}_{1}\right)}\right)+\textcolor{black}{\bm{\mu}_{t}^{\prime}\left(\mathbf{x}_{0},\mathbf{x}_{1}\right)},
\end{equation}
where $\sigma_{t}^{\prime}=\frac{d}{d t}\sigma_{t}$ and \textcolor{black}{$\bm{\mu}_{t}^{\prime}\left(\mathbf{x}_0,\mathbf{x}_1 \right)=\frac{d}{d t}\bm{\mu}_{t}\left(\mathbf{x}_{0},\mathbf{x}_{1}\right)$}. 
% Note that the standard deviation $\sigma_{t}$ is constrained to be non-zero. 
The $\bm{\mu}_t (\mathbf{x}_0,\mathbf{x}_1)$ is designed to follow the logistic schedule as
\begin{equation}\label{eq: logistic mean schedule}
	\bm{\mu}_{t} (\mathbf{x}_0,\mathbf{x}_1) = \mathbf{x}_0 + \frac{\mathbf{x}_1 - \mathbf{x}_0}{e^{k/2} - 1} \left( \frac{1 + e^{k/2}}{1 + e^{-k(t-0.5)}} - 1 \right),
\end{equation}
where $k$ determines how steeply the transition occurs. Increasing $k$ makes the switch from $\mathbf{x}_0$ to $\mathbf{x}_1$ more quickly around $t=0.5$. 
The logistic mean schedule has been shown to be able to perturb the signal more effectively \cite{wang2025targetmatchingbasedgenerative} than the widely used variance exploding, variance preserving, and linear mean schedule. 
For the variance, we employ the bridge schedule,
\begin{equation}
	\sigma_t = \sigma \sqrt{t(1-t)}
	\label{eq:bridge_variance},
\end{equation}
where $\sigma$ is a hyperparameter determining the maximal Gaussian distribution. Note that $\sigma_t$ remains non-zero and the timestep $t$ is limited between 0 and 1. 
 
\subsection{Target estimator and training objective}

Instead of estimating the whole vector field $u$ in flow-matching, we adopt the \textcolor{black}{target-matching} based generative model that directly predicts the target embedding vector $\mathbf{x}_0$. Concretely, a neural network parameterized by ${\theta}$ mapping $\mathbf{x}_t$, the conditioning $\mathbf{X}_c$, and timestep $t$ to an output $\mathbf{x}_{\theta}\left(\mathbf{x}_t,\mathbf{X}_c,t\right)$ that approximates $\mathbf{x}_0$.
Compared to the widely used score-mathcing and flow-matching based generative models, the target-matching model has been proven to be more stable and efficient \cite{wang2025targetmatchingbasedgenerative}.

The neural network $\mathbf{x}_{\theta}$ comprises three stages. In the first stage, the conditions $\mathbf{X}_c$ are weighted by learnable parameters and summed into a fused condition $\mathbf{x}_c$. In the second stage, the embedding vectors $\mathbf{x}_t$ and $\mathbf{x}_c$ are concatenated to form an embedding vector with a length of $2L$. A linear layer is then applied to project this concatenated vector into an embedding space with a dimension of $L$. This step facilitates the fusion of information between the perturbation $\mathbf{x}_t$ and the condition $\mathbf{x}_c$. In the third stage, a stacked transformer architecture is used for target estimation. Within this transformer, the timestep $t$ is explicitly injected into the network via adaptive RMS-norm  \cite{Peebles_2023_ICCV}.
\begin{table}[tbp]
	
    \refstepcounter{table}
    \label{algorithm1}
	\centering
	\renewcommand\arraystretch{1.1}
	\normalsize
	\begin{tabular}{l}
		\hline
		\textbf{Algorithm 1} Training \\
		\hline
		\quad \textbf{Input: } Data pairs $(\mathbf{s},b)$, pre-trained HuBERT   \\
		\quad For each epoch \textbf{do}:                            \\
             \quad \quad Obtain $\mathbf{x}_1$ and $\mathbf{X}_c$ according to Eq. (\ref{eq: ssl})                              \\
             \quad \quad Obtain $\mathbf{x}_0$ according to \textcolor{black}{Eq. (\ref{eq: sin})}                              \\
		\quad \quad Sample timestep $t$ from $\mathcal{U}([t_{eps},T])$                                 \\
		\quad \quad Obtain \textcolor{black}{$\bm{\mu}_{t}(\mathbf{x}_0,\mathbf{x}_1)$} and $\sigma_t$ by Eqs. (\ref{eq: logistic mean schedule}) and (\ref{eq:bridge_variance})                                 \\
		\quad \quad Sample $\mathbf{z}$ independently from $\mathcal{N}(0,1)$                              \\
		\quad \quad Obtain perturbed signal: $\mathbf{x}_t=\textcolor{black}{\bm{\mu}_{t}(\mathbf{x}_0,\mathbf{x}_1)}+\sigma_{t}\mathbf{z}$                             \\
		\quad \quad Estimate $\hat{\mathbf{x}}_{0} = \mathbf{x}_{\theta}(\mathbf{x}_t,\mathbf{X}_c,t)$                               \\
		\quad \quad Compute loss according to Eq. (\ref{eq:loss_function})                          \\
		\quad \quad Backpropagate to update $\mathbf{x}_{\theta}$                         \\
		\quad \textbf{Output:}  Optimized target predictor $\mathbf{x}_{\theta}(\mathbf{x}_t,\mathbf{X}_c,t)$\\
		\hline
		\vspace{-4mm}
	\end{tabular}
\end{table}
We use the target matching loss to train $\mathbf{x}_{\theta}$ as 
\begin{equation}
	\mathcal{L}_{tm}(\theta)=\mathbb{E}_{\mathbf{x}_t\mid \mathbf{x}_1,\mathbf{X}_c,t,\mathbf{x}_0} \left[\|\mathbf{x}_{\theta}\left(\mathbf{x}_t,\mathbf{X}_c,t\right)-\mathbf{x}_0\|^{2}\right],
	\label{eq:loss_function}
\end{equation}
where $\| \cdot \|$ denotes the Euclidean norm.

We summarize the training process in Algorithm \ref{algorithm1}, in which the target estimator $\mathbf{x}_{\theta}(\mathbf{x}_t,\mathbf{X}_c,t)$ is optimized. Given a speech signal $\mathbf{s}$ and its class label $b$, we extract the initial data sample $\mathbf{x}_1$ and the conditioning variable $\mathbf{X}_c$ using Eq. (\ref{eq: ssl}). Then we derive the target terminal sample $\mathbf{x}_0$ via \textcolor{black}{Eq. (\ref{eq: sin})}. Subsequently,
$t$ is drawn from a uniform distribution over $[\epsilon,T]$, with $\epsilon$ ensuring the  numerical stability, and $T$ being the maximum diffusion time. In the next step, the perturbed signal $\mathbf{x}_t$ is given by $	\mathbf{x}_t = \textcolor{black}{\bm{\mu}_{t}(\mathbf{x}_0,\mathbf{x}_1)} + \sigma_{t}\mathbf{z},
$ where $\mathbf{z}$ is independently sampled from $\mathcal{N}(0,1)$. \textcolor{black}{$\bm{\mu}_{t}(\mathbf{x}_0,\mathbf{x}_1)$} and $\sigma_{t}$ is calculated according to Eq. (\ref{eq: logistic mean schedule}) and  Eq. (\ref{eq:bridge_variance}), respectively. Finally, the loss described in Eq. (\ref{eq:loss_function}) is computed and backpropagated to update the parameters of $\mathbf{x}_{\theta}(\mathbf{x}_t,\mathbf{X}_c,t)$

\subsection{Sampling method}
Following the training of the target estimator, the initial sample $\hat{\mathbf{x}}_{T}$ is drawn from the speech signal. Then the vector field is calculated as
\begin{equation}
	\begin{aligned}
		\bm{u}_{\theta}(\mathbf{x}_{\theta},t,\mathbf{x}_1) &= \frac{\sigma_{t}^{\prime}}{\sigma_{t}}\left(\mathbf{x}_t-\bm{\mu}_{t}\left(\mathbf{x}_{\theta}(\mathbf{x}_t,\mathbf{X}_c,t),\mathbf{x}_1\right)\right) \\
		&\quad +\bm{\mu}_{t}^{\prime}\left(\mathbf{x}_1,\mathbf{x}_{\theta}(\mathbf{x}_t,\mathbf{X}_c,t)\right).
	\end{aligned}
	\label{eq:vector_field}
\end{equation}
The Euler ODE solver is employed to estimate terminal sample $\hat{\mathbf{x}}_0$ by iteratively computing
\begin{align}
	\hat{\mathbf{x}}_{T-n/N} &\approx \hat{\mathbf{x}}_{t}+\bm{u}_{\theta}(\hat{\mathbf{x}}_{t},t,\mathbf{x}_1)/N \label{eq:update_rule}, \\
	t &= T-n/N \label{eq:time_update},
\end{align}
where $\hat{\mathbf{x}}_t = \mathbf{x}_{\theta}(\mathbf{x}_t,\mathbf{X}_c,t)$, $N$ denotes the number of timesteps, $n$ denotes the current timestep, and $t$ is set to $T$ at the start.

% (0.97 for numerical stability in this paper). 
% The use of Euler solver provides a straightforward and efficient method for solving the ODE, ensuring stable and reproducible sampling results.
\begin{table}[tbp]
	
	\centering
    \refstepcounter{table}
	\label{algorithm2}
    \renewcommand\arraystretch{1.1}
	\normalsize
	\begin{tabular}{l}
		\hline
		\textbf{Algorithm 2} Inference \\
		\hline
		\quad \textbf{Input}: speech signal $\mathbf{s}$, target predictor $\mathbf{x}_{\theta}(\mathbf{x}_t,\mathbf{X}_c,t)$,  
		\\    \qquad \qquad number of sampling steps $N$   \\ \qquad \qquad pre-trained HuBERT \\
		\quad \textbf{Initialization: }   Obtain $\mathbf{x}_1$ and $\mathbf{X}_c$ according to Eq. (\ref{eq: ssl}), \\ \quad \quad \quad \quad $\mathbf{x}_t=\mathbf{x}_1$, $t=T$, $n=1$   \\
		\quad \textbf{while}   $n \leq N$ \textbf{do:}                         \\
		\quad \quad Compute the vector field $\bm{u}_{\theta}(\mathbf{x}_{\theta},t,\mathbf{x}_1)$ by Eq. (\ref{eq:vector_field})                             \\
		\quad \quad Update $\hat{\mathbf{x}}_{t}$ according to Eq. (\ref{eq:update_rule})                            \\
		\quad \quad Estimate $\mathbf{x}_{\theta}(\mathbf{x}_t,\mathbf{X}_c,t)$                               \\
		\quad \quad Update timestep: $t=T-n/N$, $n=n+1$ \\
	\quad \textbf{Classification:} Predict the class by Eq. (\ref{eq:sim})	 \\
        \quad \textbf{Output:}  The predicted class index $\hat{i}_b$ \\
		\hline
		\vspace{-4mm}
	\end{tabular}

\end{table}
\setcounter{table}{0} 
We calculate the cosine similarity between $\hat{\mathbf{x}}_0$ and the embedding vectors associated with each class label $b$, and the predicted class is assigned by selecting the label corresponding to the highest similarity score
\begin{equation}
    \hat{i}_b = \operatorname*{argmax}_{b} \, \frac{ \langle \hat{\mathbf{x}}_0, \mathbf{x}_0(b) \rangle }{ \| \hat{\mathbf{x}}_0 \| \, \| \mathbf{x}_0(b) \| },
    \label{eq:sim}
\end{equation}
where $ <\cdot,\cdot>$ denotes the inner product of two vectors.

In Algorithm \ref{algorithm2}, we summarize the inference process. Given a speech signal, we first extract the initial data sample $\mathbf{x}_1$ and the conditioning variable $\mathbf{X}_c$ using Eq. (\ref{eq: ssl}), and then the estimated terminal sample $\hat{\mathbf{x}}_0$ is obtained iteratively via the Euler ODE solver. During the classification stage, the predicted class is determined by the similarity between the estimated terminal sample and each class.

\section{Experimental evaluations}
\subsection{Dataset}
We mainly focus on emotion classification for English corpora and we use crema-d \cite{cao2014crema}, emodb \cite{burkhardt2005database}, TESS, savee \cite{jackson2014surrey}, RAVDESS \cite{livingstone2018ryerson}, MELD \cite{poria2018meld}, and an in-house emotion classification dataset as our training set. There are over 52k data items and 48 hours in total.  We choose MELD \cite{poria2018meld} test set to evaluate the model performance. 
For the baseline models, we choose two types of methods for comparison. The first type is deterministic model and follows the paradigm of encoder + classifier, such as emotion2vec \cite{ma2023emotion2vec} and WavLM \cite{chen2022WavLM}. The second type is a generative model and follows the paradigm of encoder + LLM, such as Qwen-audio \cite{chu2023qwen}, Qwen2-audio \cite{chu2024qwen2} and OSUM \cite{geng2025osum}.

\subsection{Implementation details}
In Eq. (\ref{eq: ssl}), we use the chinese-hubert-large\footnote{https://huggingface.co/TencentGameMate/chinese-hubert-large} model with 24 transformer layers.
In the third stage of the neural network $\mathbf{x}_{\theta}$, we employ a transformer architecture comprising four layers with an embedding dimension of 1024 and a 16-head self-attention mechanism. Combined with the neural networks from the first and second stages, the model contains a total of $71.4$ M parameters. Training is conducted with a batch size of 128 for 400k steps, using a learning rate of $5\times10^{-4}$.  We report accuracy as the evaluation metric for each model.

\subsection{Experimental results}
We compare the accuracy of the proposed method and baselines in Table \ref{SER}. The first observation is that LLM-based methods are superior to the classification methods, potentially due to the large scale of training data and the mutual reinforcement between different training tasks of LLM. In addition, the proposed method outperforms both classification methods and most LLM-based methods, which shows that the proposed method can achieve efficient distribution transport. The performance of our method is not as good as SenseVoice-L \cite{an2024funaudiollm} and potential reasons are that SenseVoice-L is trained on a large dataset comprising 30M items \cite{geng2025osum}, and it can benefit from LLM's ability to understand semantic information. 

We applied the proposed method to the gender classification task. We use an in-house dataset over 750k items, amounting to 1023 hours. The experimental results are shown in Table \ref{gender_result}. The proposed method outperforms other baselines. This shows that our method is versatile and can be extended to other speech understanding tasks.

\begin{table}[tbp]
\caption{Accuracy for SER on MELD \cite{poria2018meld} test set. CLS denotes a classification layer.}
\centering
\begin{tabular}{llc}

\toprule[1pt]
\textbf{Model} & \textbf{Model Type} & \textbf{Accuracy(\%)} \\ \hline\specialrule{0em}{1pt}{1pt}
WavLM + CLS   & Classification    & 50.6              \\ \hline\specialrule{0em}{1pt}{1pt}
Hubert + CLS   & Classification       & 53.4              \\ \hline\specialrule{0em}{1pt}{1pt}
emotion2vec \cite{ma2023emotion2vec} & Classification  & 51.9          \\ \hline\specialrule{0em}{1pt}{1pt}
Qwen-audio \cite{chu2023qwen}     &LLM     & 55.7              \\ \hline\specialrule{0em}{1pt}{1pt}
Qwen2-audio \cite{chu2024qwen2}   &LLM  & 55.3              \\ \hline\specialrule{0em}{1pt}{1pt}
OSUM \cite{geng2025osum}   &LLM   & 53.4              \\ \hline\specialrule{0em}{1pt}{1pt}
SenseVoice-L \cite{an2024funaudiollm} &LLM     & \textbf{63.1}              \\ \hline\specialrule{0em}{1pt}{1pt}
Ours  &Diffusion        &    56.5           \\ \bottomrule[1pt]
\end{tabular}
\label{SER}
\end{table}

\begin{table}[tbp]
\caption{Accuracy for gender classification on Air-Bench \cite{yang2024air}. Results of Qwen2-audio \cite{chu2024qwen2}, Qwen-audio Turbo \cite{chu2023qwen} and Soundwave \cite{zhang2025soundwave} are from \cite{zhang2025soundwave}.}
\centering
\begin{tabular}{llc}

\toprule[1pt]
\textbf{Model} & \textbf{Type} & \textbf{Accuracy(\%)} \\ \hline\specialrule{0em}{1pt}{1pt}
Fbank + CLS  & Classification       & 86.6              \\ \hline\specialrule{0em}{1pt}{1pt}
WavLM + CLS  & Classification       & 87.5              \\ \hline\specialrule{0em}{1pt}{1pt}
Qwen2-audio \cite{chu2024qwen2} & LLM   & 79.3              \\ \hline\specialrule{0em}{1pt}{1pt}
Qwen-audio Turbo \cite{chu2023qwen} & LLM        & 82.5              \\ \hline\specialrule{0em}{1pt}{1pt}
Soundwave \cite{zhang2025soundwave} & LLM    & 90.3              \\ \hline\specialrule{0em}{1pt}{1pt}
Ours   & Diffusion        & \textbf{90.5}              \\ \bottomrule[1pt]
\end{tabular}
\label{gender_result}
\end{table}

\subsection{Investigation on the distribution transport} 
\begin{figure}[!tbp]
	\centering
	\subfloat[$\mathbf{x}_1$]{
		\includegraphics[width=0.15\textwidth]{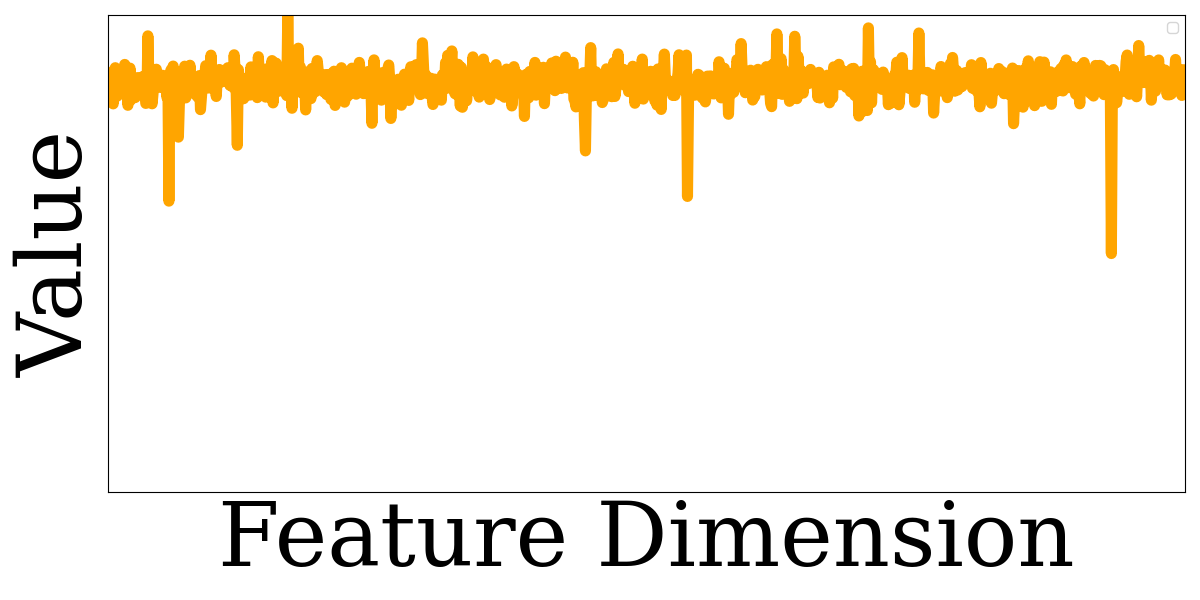}
		\label{fig:1}
	}\hfill
	\subfloat[$\hat{\mathbf{x}}_{0.75}$]{
		\includegraphics[width=0.15\textwidth]{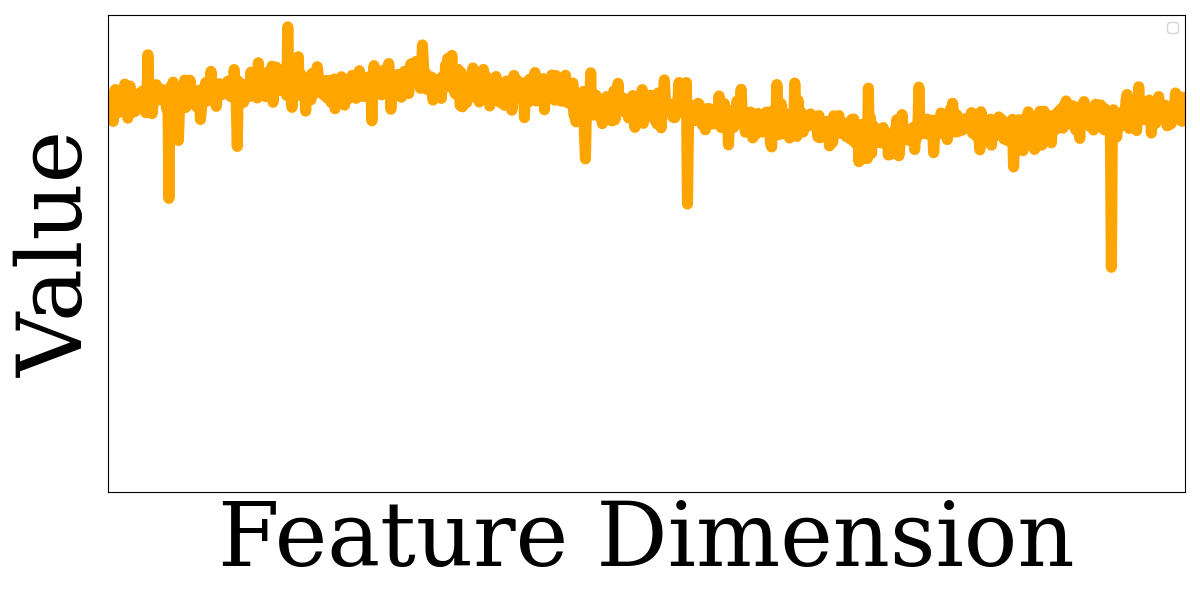}
		\label{fig:2}
	}\hfill
	\subfloat[$\hat{\mathbf{x}}_{0.5}$]{
		\includegraphics[width=0.15\textwidth]{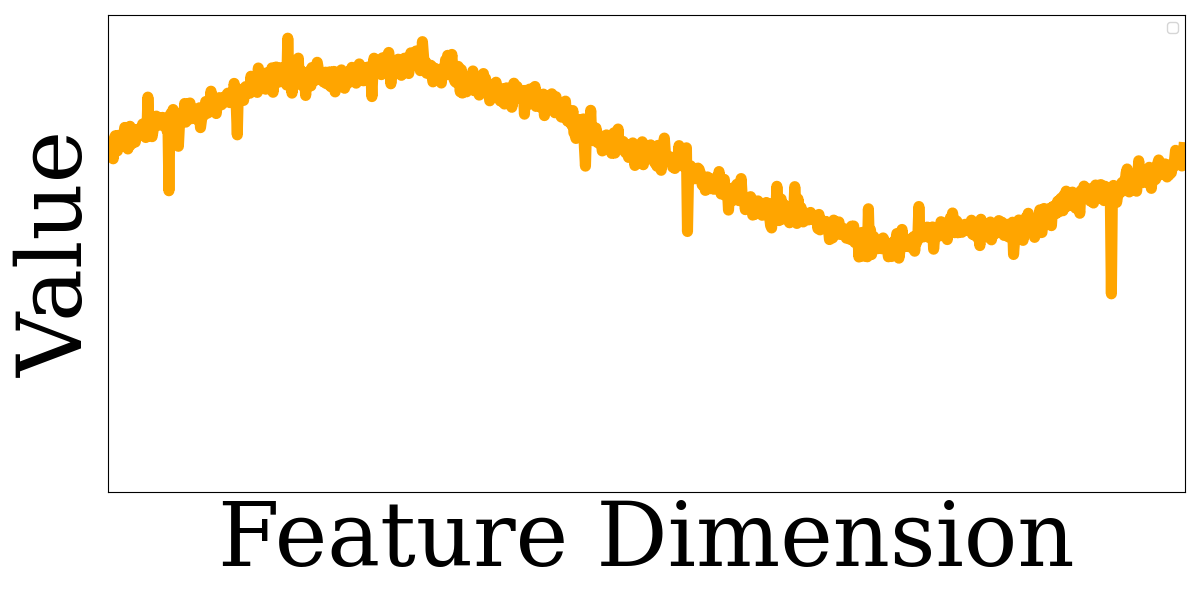}
		\label{fig:3}
	}\hfill
	\subfloat[$\hat{\mathbf{x}}_{0.25}$]{
		\includegraphics[width=0.15\textwidth]{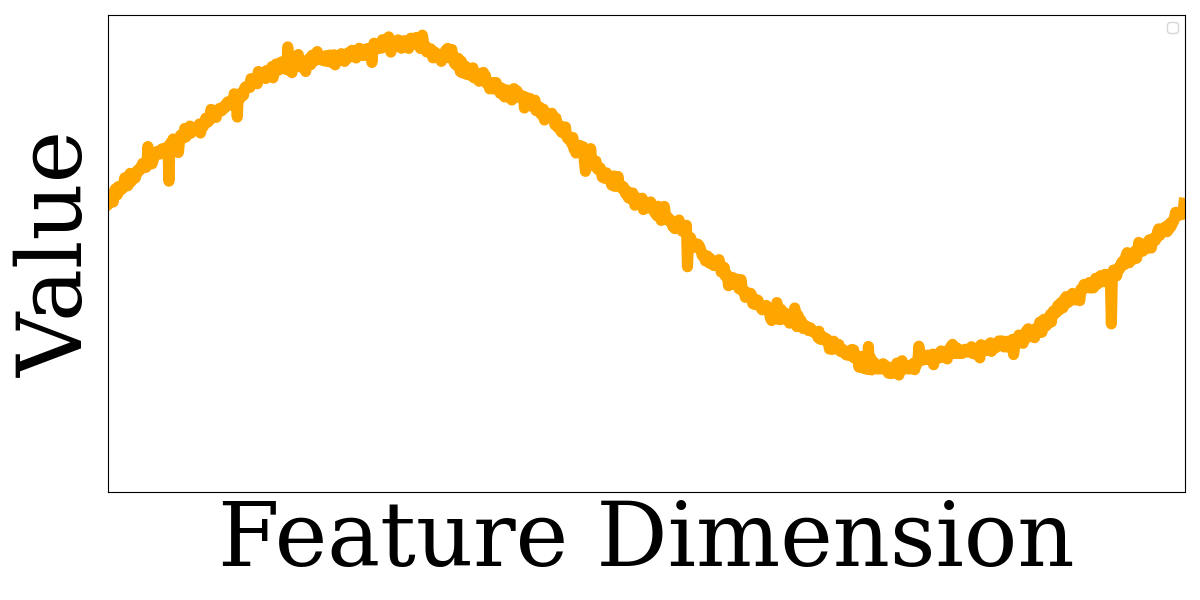}
		\label{fig:4}
	}\hfill
    	\subfloat[$\hat{\mathbf{x}}_{0.03}$]{
		\includegraphics[width=0.15\textwidth]{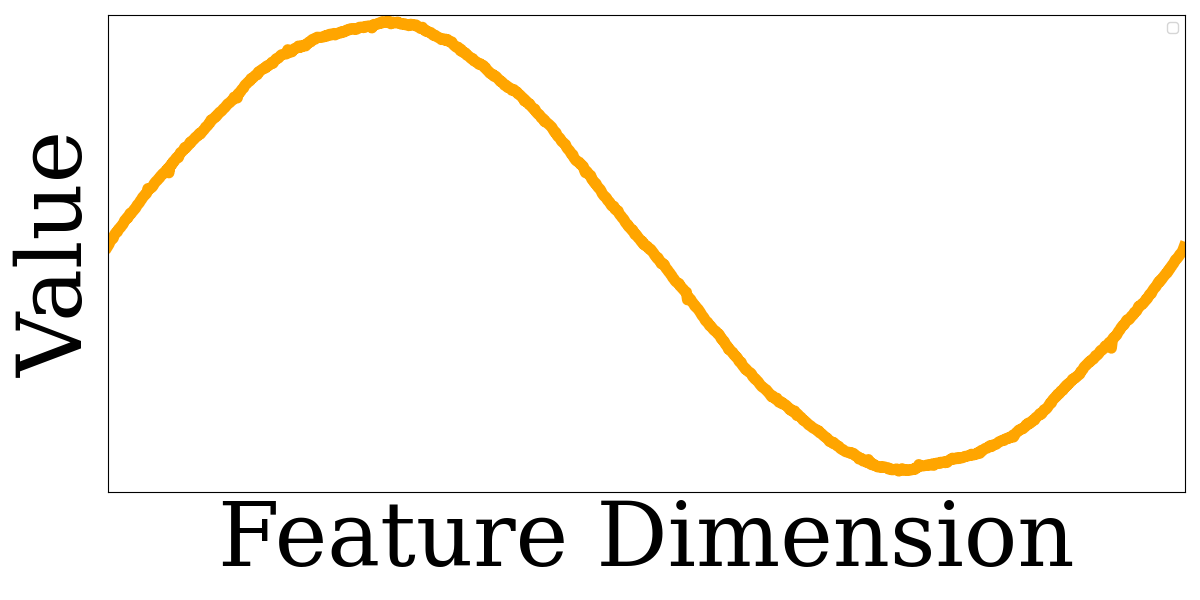}
		\label{fig:5}
	}\hfill
    	\subfloat[$\mathbf{\mathbf{x}}_0$]{
		\includegraphics[width=0.15\textwidth]{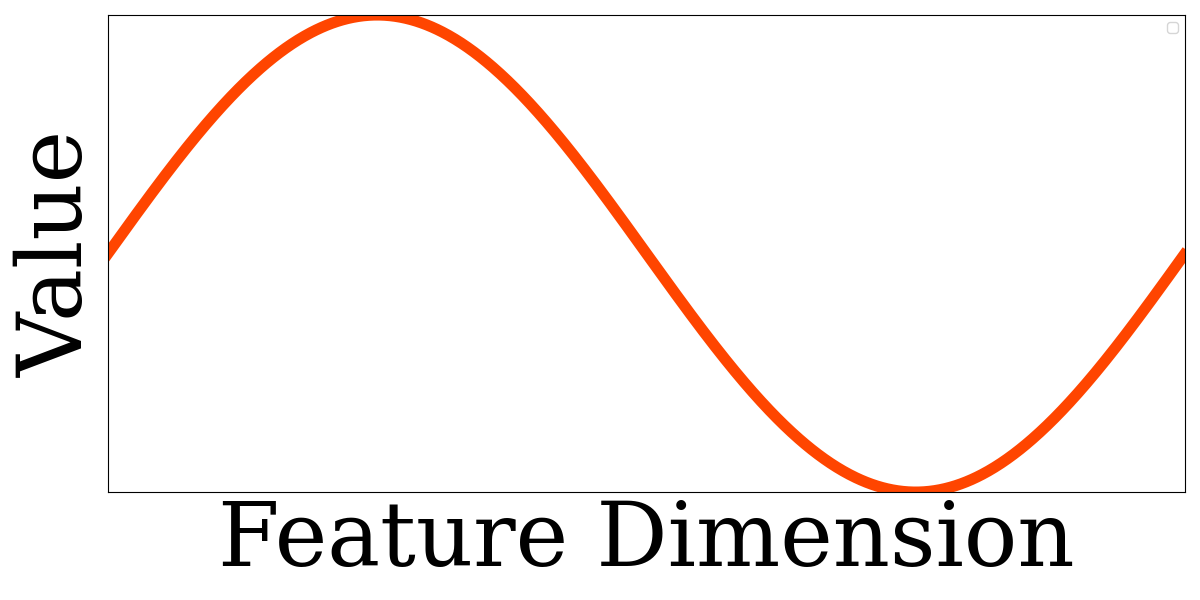}
		\label{fig:6}
	} 
	\caption{Mean transformation process at different timesteps.}
	\label{fig:distribution_trans}
\end{figure}

Fig. \ref{fig:1} shows the initial embedding vector $\mathbf{x}_1$ extracted using the Hubert model. Fig. \ref{fig:6} shows the target terminal embedding vector $\mathbf{x}_0$ generated using Eq. (\ref{eq: sin}) with label index $i_b=0$. Fig. \ref{fig:2} to Fig. \ref{fig:5} show the variation process of the estimated $\hat{\mathbf{x}}_t$ over the timestep. As the timestep decrease, we observe that the generated embedding vector $\hat{\mathbf{x}}_t$ progressively shifts from the initial embedding vector $\mathbf{x}_1$ to the target embedding vector $\mathbf{x}_0$. This result validates the effectiveness of the proposed generative distribution-shift-based classification model.
\begin{table}[!tbp]
\centering
\caption{Accuracy with different reasoning steps ($N$)}
\begin{tabular}{@{}lccccc@{}}
\toprule[1pt]
$N$    & 1      & 2           & 4            & 10     & 20     \\ \hline\specialrule{0em}{1pt}{1pt}
\textbf{Accuracy} & 0.5613 & 0.5617 & 0.5625  & 0.5628  & 0.5644 \\ \bottomrule[1pt]
\end{tabular}

\label{tab:accuracy_steps}
\end{table}
One issue with generative models is their reasoning steps, and hence we investigate the impact of the number of reasoning steps on the proposed method. 
Table \ref{tab:accuracy_steps} illustrates the performance of the proposed method with respect to the number of reasoning steps. We observed that the number of reasoning steps steadily improves the accuracy, and the performance of single-step reasoning is already satisfactory. 
% This is attributed to the TM-based generative model, which has been proven to outperform traditional diffusion-based and flow matching-based generative models in terms of training and inference efficiency.

\section{Conclusion}
We have presented Gen-SER, a generative framework that recasts speech emotion recognition as distribution transport: discrete emotion labels are mapped into continuous embedding vectors via sinusoidal taxonomy encoding to serve as the diffusion targets for HuBERT speech features.
The target-matching-based generative model with logistic mean and bridge variance schedules is adopted to realize efficient distribution transport. Experimental results demonstrate that the proposed method shows competitive performance in SER and is applicable in other tasks, demonstrating a new route for speech understanding.

% \vfill\pagebreak

% \section{REFERENCES}
% \label{sec:refs}

% List and number all bibliographical references at the end of the
% paper. The references can be numbered in alphabetic order or in
% order of appearance in the document. When referring to them in
% the text, type the corresponding reference number in square
% brackets as shown at the end of this sentence \cite{C2}. An
% additional final page (the fifth page, in most cases) is
% allowed, but must contain only references to the prior
% literature.

% Please follow the IEEE Citation Guidelines, \url{https://ieee-dataport.org/sites/default/files/analysis/27/IEEE\%20Citation\%20Guidelines.pdf} for formatting of references.

{
\small
\bibliographystyle{IEEEbib}
\bibliography{strings,refs}

@InProceedings{Peebles_2023_ICCV,
    author    = {Peebles, William and Xie, Saining},
    title     = {Scalable Diffusion Models with Transformers},
    booktitle = {Proceedings of the IEEE/CVF International Conference on Computer Vision (ICCV)},
    month     = {October},
    year      = {2023},
    pages     = {4195-4205}
}

@article{wang2025targetmatchingbasedgenerative,
  title={Target matching based generative model for speech enhancement},
  author={Wang, Taihui and Chen, Rilin and Lei, Tong and others},
  journal={arXiv preprint arXiv:2509.07521},
  year={2025}
}

@article{an2024funaudiollm,
  title={Funaudiollm: Voice understanding and generation foundation models for natural interaction between humans and llms},
  author={An, Keyu and Chen, Qian and Deng, Chong and others},
  journal={arXiv preprint arXiv:2407.04051},
  year={2024}
}

@inproceedings{lipmanFlowMatchingGenerative2023,
	title = {Flow Matching for Generative Modeling},
	author = {Lipman, Yaron and Chen, Ricky T. Q. and Ben-Hamu, Heli and Nickel, Maximilian and Le, Matt},
	booktitle={Proc. ICLR},
	date = {2023-02-08},
	year = {2023},
	month = Feb,
	eprint = {2210.02747},
	eprinttype = {arXiv},
	eprintclass = {cs},
	doi = {10.48550/arXiv.2210.02747},
	urldate = {2025-03-28},
	pubstate = {prepublished}
}

@inproceedings{chenFlowTSVADTargetSpeakerVoice2025,
  title={Flow-TSVAD: Target-Speaker Voice Activity Detection via Latent Flow Matching for Speaker Diarization},
  author={Chen, Zhengyang and Han, Bing and Wang, Shuai and others},
  booktitle={International Conference on Acoustics, Speech and Signal Processing (ICASSP)},
  pages={1--5},
  year={2025},
  organization={IEEE}
}

@article{ma2023emotion2vec,
  title={emotion2vec: Self-supervised pre-training for speech emotion representation},
  author={Ma, Ziyang and Zheng, Zhisheng and Ye, Jiaxin and others},
  journal={arXiv preprint arXiv:2312.15185},
  year={2023}
}

@article{mustaqeem2019cnn,
  title={A CNN-assisted enhanced audio signal processing for speech emotion recognition},
  author={Mustaqeem, N and Kwon, Soonil},
  journal={Sensors},
  volume={20},
  number={1},
  pages={183},
  year={2019},
  publisher={MDPI}
}

@article{lieskovska2021review,
  title={A review on speech emotion recognition using deep learning and attention mechanism},
  author={Lieskovsk{\'a}, Eva and Jakubec, Maro{\v{s}} and Jarina, Roman and Chmul{\'\i}k, Michal},
  journal={Electronics},
  volume={10},
  number={10},
  pages={1163},
  year={2021},
  publisher={MDPI}
}

@inproceedings{huahu2010application,
  title={Application of speech emotion recognition in intelligent household robot},
  author={Xu, Huahu and Gao, Jue and Yuan, Jian},
  booktitle={2010 International conference on artificial intelligence and computational intelligence},
  volume={1},
  pages={537--541},
  year={2010},
  organization={IEEE}
}

@article{geng2025osum,
  title={OSUM: Advancing open speech understanding models with limited resources in academia},
  author={Geng, Xuelong and Wei, Kun and Shao, Qijie and others},
  journal={arXiv preprint arXiv:2501.13306},
  year={2025}
}

@article{cao2014crema,
  title={Crema-d: Crowd-sourced emotional multimodal actors dataset},
  author={Cao, Houwei and Cooper, David G and Keutmann, Michael K and others},
  journal={IEEE transactions on affective computing},
  volume={5},
  number={4},
  pages={377--390},
  year={2014},
  publisher={IEEE}
}

@inproceedings{burkhardt2005database,
  title={A database of german emotional speech.},
  author={Burkhardt, Felix and Paeschke, Astrid and Rolfes, Miriam and others},
  booktitle={Interspeech},
  volume={5},
  pages={1517--1520},
  year={2005}
}

@article{jackson2014surrey,
  title={Surrey audio-visual expressed emotion (savee) database},
  author={Jackson, Philip and Haq, Sjuosg},
  journal={University of Surrey: Guildford, UK},
  year={2014}
}

@article{livingstone2018ryerson,
  title={The Ryerson Audio-Visual Database of Emotional Speech and Song (RAVDESS): A dynamic, multimodal set of facial and vocal expressions in North American English},
  author={Livingstone, Steven R and Russo, Frank A},
  journal={PloS one},
  volume={13},
  number={5},
  pages={e0196391},
  year={2018},
  publisher={Public Library of Science San Francisco, CA USA}
}

@article{poria2018meld,
  title={Meld: A multimodal multi-party dataset for emotion recognition in conversations},
  author={Poria, Soujanya and Hazarika, Devamanyu and Majumder, Navonil and others},
  journal={arXiv preprint arXiv:1810.02508},
  year={2018}
}

@inproceedings{pandey2019deep,
  title={Deep learning techniques for speech emotion recognition: A review},
  author={Pandey, Sandeep Kumar and Shekhawat, Hanumant Singh and Prasanna, SR Mahadeva},
  booktitle={2019 29th international conference RADIOELEKTRONIKA},
  pages={1--6},
  year={2019},
  organization={IEEE}
}

@article{du2025cosyvoice,
  title={Cosyvoice 3: Towards in-the-wild speech generation via scaling-up and post-training},
  author={Du, Zhihao and Gao, Changfeng and Wang, Yuxuan and others},
  journal={arXiv preprint arXiv:2505.17589},
  year={2025}
}

@article{gao2025differentiable,
  title={Differentiable Reward Optimization for LLM based TTS system},
  author={Gao, Changfeng and Du, Zhihao and Zhang, Shiliang},
  journal={arXiv preprint arXiv:2507.05911},
  year={2025}
}

@article{zhang2025soundwave,
  title={Soundwave: Less is More for Speech-Text Alignment in LLMs},
  author={Zhang, Yuhao and Liu, Zhiheng and Bu, Fan and others},
  journal={arXiv preprint arXiv:2502.12900},
  year={2025}
}

@article{yang2024air,
  title={Air-bench: Benchmarking large audio-language models via generative comprehension},
  author={Yang, Qian and Xu, Jin and Liu, Wenrui and others},
  journal={arXiv preprint arXiv:2402.07729},
  year={2024}
}

@article{chu2024qwen2,
  title={Qwen2-audio technical report},
  author={Chu, Yunfei and Xu, Jin and Yang, Qian and others},
  journal={arXiv preprint arXiv:2407.10759},
  year={2024}
}

@article{chu2023qwen,
  title={Qwen-audio: Advancing universal audio understanding via unified large-scale audio-language models},
  author={Chu, Yunfei and Xu, Jin and Zhou, Xiaohuan and others},
  journal={arXiv preprint arXiv:2311.07919},
  year={2023}
}

@article{chen2022wavlm,
  title={Wavlm: Large-scale self-supervised pre-training for full stack speech processing},
  author={Chen, Sanyuan and Wang, Chengyi and Chen, Zhengyang and others},
  journal={IEEE Journal of Selected Topics in Signal Processing},
  volume={16},
  number={6},
  pages={1505--1518},
  year={2022},
  publisher={IEEE}
}

@article{hsuHuBERTSelfSupervisedSpeech2021,
  title={Hubert: Self-supervised speech representation learning by masked prediction of hidden units},
  author={Hsu, Wei-Ning and Bolte, Benjamin and Tsai, Yao-Hung Hubert and others},
  journal={IEEE/ACM transactions on audio, speech, and language processing},
  volume={29},
  pages={3451--3460},
  year={2021},
  publisher={IEEE}
}

@inproceedings{liYourDiffusionModel2023,
  title={Your diffusion model is secretly a zero-shot classifier},
  author={Li, Alexander C and Prabhudesai, Mihir and Duggal, Shivam and others},
  booktitle={Proceedings of the IEEE/CVF International Conference on Computer Vision},
  pages={2206--2217},
  year={2023}
}
}
\end{document}